\documentclass[useAMS,usenatbib]{mn2e}
\usepackage{graphicx}

\newcommand{\teff}{${T}_{\mathrm{eff}}$}
\newcommand{\logg}{$\log{g}$}
\newcommand{\msun}{$M_{\odot}$}

\newcommand{\kms}{km s$^{-1}$}
\newcommand{\muhz}{$\mu$Hz}

\def\aap{A\&A}
\def\apjl{ApJ}
\def\apj{ApJ}
\def\apjs{ApJS}
\def\aj{AJ}
\def\mnras{MNRAS}

\def\pasp{PASP}

\def\aaps{A\&AS}
\def\araa{ARA\&A}

\title[The fourth and fifth pulsating ELM WDs]{A new class of pulsating white dwarf of extremely low mass: the fourth and fifth members}

\author[J. J. Hermes et al.]{J.~J.~Hermes$^{1,2}$\thanks{jjhermes@astro.as.utexas.edu},
M. H. Montgomery$^{1}$,
A. Gianninas$^{3}$,
D. E. Winget$^{1}$,
\newauthor
Warren R. Brown$^{4}$,
Samuel T. Harrold$^{1}$,
Keaton J. Bell$^{1}$,
Scott J. Kenyon$^{4}$,
\newauthor
Mukremin Kilic$^{3}$, and
Barbara G. Castanheira$^{1}$
\\
$^{1}$Department of Astronomy, University of Texas at Austin, Austin, TX\,-\,78712, USA\\
$^{2}$Department of Physics, University of Warwick, Coventry CV4 7AL, United Kingdom\\
$^{3}$Homer L. Dodge Department of Physics and Astronomy, University of Oklahoma, 440 W. Brooks St., Norman, OK\,-\,73019, USA\\
$^{4}$Smithsonian Astrophysical Observatory, 60 Garden St, Cambridge, MA\,-\,02138, USA}

\begin{document}

\maketitle

\label{firstpage}

\begin{abstract}

We report the discovery of two new pulsating extremely low-mass (ELM) white dwarfs (WDs), SDSS~J161431.28+191219.4 (hereafter J1614) and SDSS~J222859.93+362359.6 (hereafter J2228). Both WDs have masses $<$ 0.25 \msun\ and thus likely harbor helium cores. Spectral fits indicate these are the two coolest pulsating WDs ever found. J1614 has \teff\ $=8880\pm170$ K and \logg\ $=6.66\pm0.14$, which corresponds to a $\sim$0.19 \msun\ WD. J2228 is considerably cooler, with a \teff\ $=7870\pm120$ K and \logg\ $=6.03\pm0.08$, which corresponds to a $\sim$0.16 \msun\ WD, making it the coolest and lowest-mass pulsating WD known. There are multiple ELM WDs with effective temperatures between the warmest and coolest known ELM pulsators that do not pulsate to observable amplitudes, which questions the purity of the instability strip for low-mass WDs. In contrast to the CO-core ZZ Ceti stars, which are believed to represent a stage in the evolution of all such WDs, ELM WDs may not all evolve as a simple cooling sequence through an instability strip. Both stars exhibit long-period variability ($1184-6235$ s) consistent with non-radial $g$-mode pulsations. Although ELM WDs are preferentially found in close binary systems, both J1614 and J2228 do not exhibit significant radial-velocity variability, and are perhaps in low-inclination systems or have low-mass companions. These are the fourth and fifth pulsating ELM WDs known, all of which have hydrogen-dominated atmospheres, establishing these objects as a new class of pulsating WD.

\end{abstract}


\section{Introduction}

White dwarf (WD) stars represent the end points of stellar evolution for all low-mass stars, and are the fate of more than 97\% of all stars in our Galaxy. Roughly 80\% of WDs belong to the spectral class DA, with atmospheres characteristically dominated by hydrogen \citep{Kleinman13}. When DA WDs cool to the appropriate temperature to foster a hydrogen partial ionization zone, they begin their journey through the ZZ Ceti (or DAV) instability strip, where global pulsations are driven to observable amplitudes and their fundamental parameters can be determined using asteroseismology (see reviews by \citealt{WinKep08}, \citealt{FontBrass08} and \citealt{Althaus10}).

Aside from their variability, the ZZ Ceti stars discovered to date appear to be otherwise normal WDs, and are therefore believed to be a natural phase in the evolution of all DAs. Although some DAs within the empirical instability strip have been observed not to vary to modest limits (e.g., \citealt{KeplerNelan93,Mukadam04}), follow-up observations have shown that some of these stars really do pulsate at low amplitude (e.g., \citealt{Castanheira07}). Higher-quality optical and UV spectra have also moved some of these non-variable interlopers out of the instability strip \citep{Bergeron95}. Thus, it is currently believed that the ZZ Ceti instability strip is pure, and that all DA WDs will at some point pass through it and pulsate \citep{Fontaine82,Fontaine85,Bergeron04}.

Much work has been devoted to observationally mapping the ZZ Ceti instability strip, which runs in temperature from roughly $12{,}600-11{,}100$ K for standard \logg\ = 8.0 WDs \citep{Mukadam04,Gianninas11}. There is also a dependence on surface gravity, such that WDs with lower \logg\ pulsate at lower effective temperatures. This trend has been observed for WDs with masses from 1.1 \msun\ down to 0.5 \msun\ \citep{Giovannini98}.

The blue edge of the ZZ Ceti instability strip, where pulsations are turning on, has been successfully estimated by both convective period arguments \citep{Brickhill91} and full non-adiabatic calculations \citep{Winget82,Brassard99,VanGrootel12}. A slightly more efficient prescription for convection has to be assumed, by increasing the value of the mixing-length theory parameter ML2/$\alpha$, to make the theory match the observed blue edge, which was most recently mapped empirically by \citet{Gianninas11}.

However, estimating the temperature at which pulsations should shut down has remained a challenge. Modern non-adiabatic calculations do not predict a red edge until around 5600 K \citep{VanGrootel12}, more than 5000 K cooler than the empirical red edge \citep{Kanaan02,Gianninas11}. \citet{Hansen85} argue that a surface reflection criterion can be enforced to limit the maximum mode period, which may push a theoretical red edge to hotter temperatures, nearer what is observed in ZZ Ceti stars \citep{VanGrootel13}.

The recent discovery of pulsating extremely low-mass (ELM, $\leq$ 0.25 \msun) WDs provides us with an exciting new opportunity to explore the nature of the physics of WD pulsations at cooler temperatures and much lower masses. Since the first discovery by \citet{Landolt68}, more than 160 ZZ Ceti stars have been found, all of which have masses $\geq$ 0.5 \msun\ and thus likely harbor carbon-oxygen (CO) cores. That changed with the discovery of the first three pulsating ELM WDs \citep{Hermes12,Hermes13}. These ELM WDs are likely the product of binary evolution, since the Galaxy is not old enough to produce such low-mass WDs through single-star evolution \citep{Marsh95}. During a common-envelope phase, the ELM WDs were most likely stripped of enough mass to prevent helium ignition, suggesting they harbor He cores.

The pulsating ELM WDs will be incredibly useful in constraining the interior composition, hydrogen-layer mass, overall mass, rotation rate, and the behavior of convection in these low-mass WDs, which may derive a majority of their luminosities from stable hydrogen burning for the lowest-mass objects \citep{Steinfadt10}. Several groups have recently investigated the pulsation properties of He-core WDs, and non-adiabatic calculations have shown that non-radial $p$- and $g$-modes should be unstable and thus observable in these objects \citep{Corsico12,VanGrootel13}. Pulsating ELM WDs will also extend our empirical studies of the ZZ Ceti instability strip to significantly lower surface gravities.

Boosted by the many new ELM WDs catalogued by the ELM Survey, a targeted spectroscopic search for ELM WDs \citep{BrownELMiii,KilicELMiv,BrownELMv}, we have looked for additional pulsating ELM WDs throughout a large part of parameter space. The first three pulsating ELM WDs all have effective temperatures below $10{,}000$ K, much cooler than any previously known CO-core ZZ Ceti star \citep{Mukadam04}, which makes up the coolest class of pulsating WDs. We now add to that list the two coolest pulsating WDs ever found, SDSS~J161431.28+191219.4 ($g=16.4$ mag, hereafter J1614) and SDSS~J222859.93+362359.6 ($g=16.9$ mag, hereafter J2228), bringing to five the number of ELM WDs known to pulsate.

In Section~\ref{sec:J1614} we detail our discovery of pulsations in J1614 and outline our new spectroscopic observations of this ELM WD. In Section~\ref{sec:J2228} we describe the discovery of multi-periodic variability in the ELM WD J2228 and update its determined atmospheric parameters. We conclude with a discussion of these discoveries, and update the observed DA WD instability strip.



\section{SDSS J161431.28+191219.4}
\label{sec:J1614}

\subsection{Spectroscopic Observations}
\label{sec:J1614spec}

\begin{figure}
\centering{\includegraphics[width=0.85\columnwidth]{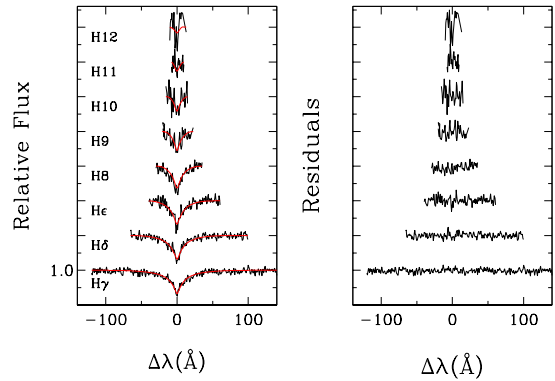}}
\caption{Model atmosphere fits (red) to the observed hydrogen Balmer lines (black) from H$\gamma$$-$H12 for the summed spectra of J1614 taken from the FLWO 1.5 m telescope. This model derives the primary parameters in Section~\ref{sec:J1614atm}. The individual Balmer lines are normalized to unity and offset vertically by a factor of 0.3 for clarity. Residuals from the model fit are shown at right. \label{fig:J1614spec}}
\end{figure}

\citet{BrownELMiii} found that J1614 had \teff\ $=8590\pm540$ K and \logg\ $= 5.64\pm0.12$, based on a single spectrum of this $g=16.4$ mag WD from the FLWO 1.5 m telescope using the FAST spectrograph \citep{Fabricant98}. We have obtained an additional 51 spectra using the same instrument and setup.

\subsubsection{Atmospheric Parameters}
\label{sec:J1614atm}

We have co-added our spectroscopic observations to determine the atmospheric parameters of the ELM WD J1614 (Figure~\ref{fig:J1614spec}). Our observations cover a wavelength range from $3550-5450$ \AA. The model atmospheres used for this analysis are described at length in \citet{Gianninas11} and employ the new Stark broadening profiles from \citet{TB09}. Models where convective energy transport becomes important are computed using the ML2/$\alpha$ = 0.8 prescription of the mixing-length theory \citep[see][]{tremblay10}. A discussion of our extension of these models to lower surface gravities and more details of our fitting method can be found in Section 2.1.1 of \citet{Hermes13}.

Our final fit to the phased and co-added spectrum of J1614 is shown in the top panel of Figure~\ref{fig:J1614spec} and yields \teff\ $=8880\pm170$ K and \logg\ $=6.66\pm0.14$. This corresponds to a mass of $\sim$0.20 \msun\ using the He-core WD models of \citet{Panei07}, if we assume the WD is in its final cooling stage. The more recent models of \citet{Althaus13} predict a mass of 0.19 \msun\ given the atmospheric parameters, which we adopt.

We have also performed our fit without using the low S/N lines H11$-$H12, but this marginally affects our solution: Using only the H$\gamma$$-$H10 lines of the Balmer series, we find \teff\ $=8830\pm160$ K and \logg\ $=6.54\pm0.16$. To remain consistent with our previous pulsating ELM WD atmospheric determinations \citep{Hermes12,Hermes13}, we will include the H11$-$H12 lines in our adopted solution for J1614.

\begin{table}
\begin{center}
 \centering
  \caption{Journal of photometric observations.}\label{tab:jour}
  \begin{tabular}{@{}lcccc@{}}
  \hline
Run & UT Date & Length & Seeing & Exp. \\
  &   & (hr) & (\arcsec) & (s) \\
 \hline
\multicolumn{5}{c}{\bf SDSS J161431.28+191219.4} \\
A2690	&	2012 Jun 21	&	2.6	&	1.7	&	5	\\
A2692	&	2012 Jun 22	&	2.0	&	1.8	&	5	\\
A2695	&	2012 Jun 23	&	3.7	&	1.2	&	5	\\
A2697	&	2012 Jun 24	&	3.5	&	1.4	&	5	\\
A2699	&	2012 Jun 25	&	3.6	&	1.3	&	5	\\
\hline
\multicolumn{5}{c}{\bf SDSS J222859.93+362359.6} \\
A2521	&	2011 Nov 28	&	3.5	&	1.5	&	10	\\
A2524	&	2011 Nov 29	&	1.6	&	2.5	&	10	\\
A2528	&	2011 Nov 30	&	1.9	&	2.2	&	10	\\
A2707	&	2012 Jul 13	&	2.3	&	1.1	&	5	\\
A2710	&	2012 Sep 17	&	2.8	&	2.4	&	10	\\
A2719	&	2012 Sep 20	&	6.4	&	1.6	&	15	\\
A2721	&	2012 Sep 21	&	7.4	&	1.4	&	10	\\
\hline
\end{tabular}
\end{center}
\end{table}

\subsubsection{Radial Velocity Observations}
\label{sec:J1614rv}

ELM WDs are typically found in close binary systems; these companions are necessary to strip the progenitor of enough mass to form such a low-mass WD within the age of the Universe \citep{KilicELMii}. However, using the code of \citet{Kenyon86}, we do not detect any significant radial velocity variability in our observations of J1614. The r.m.s. scatter gives us an upper limit on the RV semi-amplitude: $K<56$ \kms. The systemic velocity is $\gamma=-148.7\pm7.6$ \kms.

We note that this non-detection does not require the lack of a companion to the ELM WD in J1614. Rather, the system may be inclined nearly face-on to our line of sight, or the companion may be a much cooler low-mass WD. If the inclination is $i>30 \degr$, which is more than 85\% likely if the orientation of the system with respect to the Earth is drawn from a random distribution, the companion has $M_2<0.17$ \msun\ if the system has a 7 hr orbital period, the median for ELM WD binaries in the ELM Survey \citep{BrownELMv}. Empirically, there are similarly low-mass WDs in the ELM Survey with no significant radial velocity variability \citep{BrownELMiii}.

\subsection{Photometric Observations}
\label{sec:J1614photo}

\begin{figure}
\centering{\includegraphics[width=0.99\columnwidth]{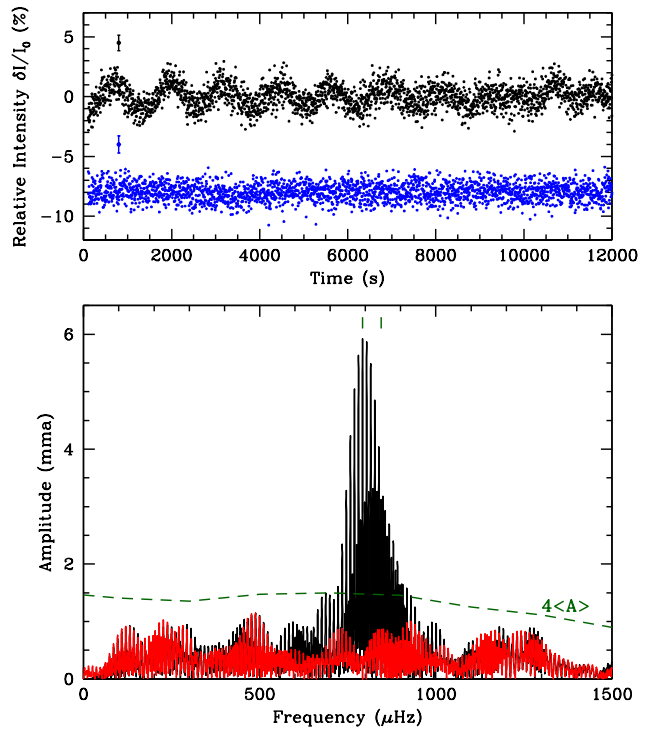}}
\caption{The top panel shows high-speed photometry of J1614 from a representative run on 2012 June 23. The brightest comparison star is shown in blue, offset by $-8$\%. Average point-by-point photometric errors are also shown. The bottom panel shows a Fourier transform of our entire data set to date, some 15.4 hr of observations in 2012~June. We also display in red the FT of the residuals after prewhitening by the two periods listed in Table~\ref{tab:J1614freq}, and mark those periods with green tick marks at the top of the panel. We mark the 4$\langle {\rm A}\rangle$ significance level, described in the text, as a dashed green line. \label{fig:J1614}}
\end{figure}

We obtained high-speed photometric observations of J1614 at the McDonald Observatory over five consecutive nights in 2012~June for a total of nearly 15.4 hr of coverage. We used the Argos instrument, a frame-transfer CCD mounted at the prime focus of the 2.1m Otto Struve telescope \citep{Nather04}, to obtain $5$ s exposures on J1614. A full journal of observations can be found in Table~\ref{tab:jour}. Observations were obtained through a 3mm BG40 filter to reduce sky noise.

We performed weighted, circular, aperture photometry on the calibrated frames using the external IRAF package $\textit{ccd\_hsp}$ written by Antonio Kanaan \citep{Kanaan02}. We divided the sky-subtracted light curves by the brightest comparison star in the field, SDSS~J161433.39+191058.3 ($g=14.3$ mag), to correct for transparency variations, and applied a timing correction to each observation to account for the motion of the Earth around the barycenter of the solar system \citep{Stumpff80,Thompson09}.

The top panel of Figure~\ref{fig:J1614} shows a portion of a typical light curve for J1614, obtained on 2012 June 23, and includes the brightest comparison star in the field over the same period. The bottom panel of this figure shows a Fourier transform (FT) utilizing all $11{,}075$ light curve points collected thus far. We display the 4$\langle {\rm A}\rangle$ significance line at the bottom of Figure~\ref{fig:J1614}, calculated from the average amplitude, $\langle {\rm A}\rangle$, of an FT within a 1000 \muhz\ region in steps of 200 \muhz, after pre-whitening by the two highest-amplitude periodicities.

\begin{table}
 \centering
  \caption{Frequency solution for SDSS J161431.28+191219.4 \label{tab:J1614freq}}
  \begin{tabular}{@{}lcccc@{}}
  \hline
  ID & Period & Frequency & Amplitude & S/N \\
   & (s) & ($\mu$Hz) & (mma) &  \\
 \hline
$f_1$ & $1262.668\pm0.041$ & $791.974\pm0.026$ & $5.94\pm0.11$ & 16.0 \\
$f_2$ & $1184.106\pm0.064$ & $844.519\pm0.045$ & $3.20\pm0.10$ & 8.6 \\
\hline
\end{tabular}
\end{table}

The pulse shape of J1614 appears quite sinusoidal, and is well described by two nearby periods at 1262.67 and 1184.11 s. The amplitudes of these periods are identified in Table~\ref{tab:J1614freq}, where 1 mma = 0.1\% relative amplitude. For more realistic estimates, the cited errors are not formal least-squares errors to the data but rather the product of $10^5$ Monte Carlo simulations of perturbed data using the software package Period04 \citep{Lenz05}. The signal-to-noise calculation is based on the amplitude of the variability as compared to the average amplitude of a 1000 \muhz\ box centered around that variability, after pre-whitening by the two highest-amplitude periodicities.



\section{SDSS J222859.93+362359.6}
\label{sec:J2228}

\subsection{Spectroscopic Observations}
\label{sec:J2228spec}

We targeted J2228 based on a single classification spectrum published in \citet{BrownELMiii}. A preliminary fit to the spectrum of this $g=16.9$ mag WD from the FLWO 1.5 m telescope using the FAST spectrograph found \teff\ $=8590\pm540$ K and \logg\ $= 5.64\pm0.12$. We have obtained 30 additional spectra using the FLWO 1.5 m telescope and two additional spectra using the Blue Channel Spectrograph \citep{Schmidt89} on the 6.5m MMT.

\subsubsection{Atmospheric Parameters}
\label{sec:J2228atm}

As with J1614, we have co-added our spectroscopic observations to determine the atmospheric parameters of the primary ELM WD visible in J2228. Our model atmosphere fitting is identical to that as described in Section~\ref{sec:J1614atm}.

\begin{figure}
\centering{\includegraphics[width=0.85\columnwidth]{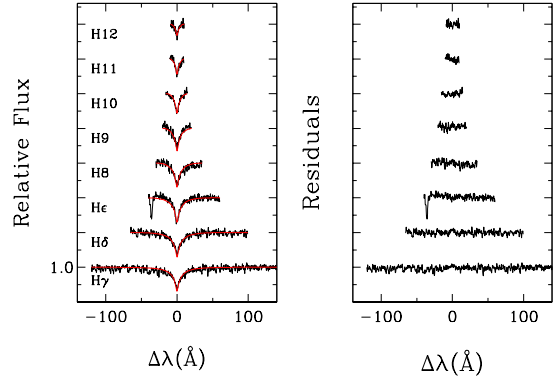}}
\caption{Model atmosphere fits (red) to the observed hydrogen Balmer lines (black) from H$\gamma$$-$H12 for the summed spectra of J2228 taken from the MMT. This model derives the primary parameters in Section~\ref{sec:J2228atm}. The individual Balmer lines are normalized to unity and offset vertically by a factor of 0.3 for clarity. Residuals from the model fit are shown at right. Note that we exclude the spectral range containing the observed Ca K line, from 3925 to 3945 \AA\ in the blue wing of the H$\epsilon$ line, from both the normalization procedure and the fitting routine. \label{fig:J2228spec}}
\end{figure}

Our final fit to the co-added MMT spectrum of J2228 is shown in Figure~\ref{fig:J2228spec} and yields \teff\ $=7870\pm120$ K and \logg\ $=6.03\pm0.08$. This corresponds to a mass of $\sim$0.16 \msun\ using the He-core WD models of both \citet{Panei07} and \citet{Althaus13}. Similarly, a fit to our 31 FAST spectra, which has lower S/N, finds \teff\ $=7990\pm190$ K and \logg\ $=6.25\pm0.15$.

In addition to the Balmer series, the Ca II K line is also observed in the spectra of J2228, seen in absorption in the blue wing of the H$\epsilon$ line seen in Figure~\ref{fig:J2228spec}. Strong Ca lines have been seen before in very low-surface-gravity WDs (\logg\ $< 6.0$), and these metal lines typically phase with the ELM WD radial velocity and are thus not interstellar (e.g. \citealt{Hermes13,BrownELMv,Kaplan13}). For the purposes of this analysis, we simply exclude the wavelength range where this metal line is present so that it does not affect either the normalization of the individual Balmer lines nor the actual fits themselves. The presence of Ca in the photosphere of this WD should not introduce a systematic effect on the derived atmospheric parameters \citep{Gianninas04}.

\subsubsection{Radial Velocity Observations}
\label{sec:J2228rv}

As with J1614, we do not detect any significant radial velocity variability in our observations of J2228. Again, the r.m.s. scatter gives us an upper limit on the RV semi-amplitude: $K<28$ \kms. The systemic velocity is $\gamma=-52.5\pm4.7$ \kms.

If the inclination is $i>30 \degr$, we can put an upper limit on the companion mass of $M_2<0.06$ \msun\ if the orbital period is 7 hr (see Section~\ref{sec:J1614rv}). Similarly, the companion would have less mass than $M_2<0.26$ \msun\ if the inclination is $i>10 \degr$ (there is a $< 2$\% probability that the inclination is $i<10 \degr$ if the system is a member of a group of stars whose orientations with respect to the Earth are random). The ELM WD we observe required a companion in order to to lose enough mass to get to its present configuration, so unless that companion was ejected from the binary, it likely has a very low mass.

\subsection{Photometric Observations}
\label{sec:J2228photo}

Our high-speed photometric observations of J2228 were obtained and reduced in an identical manner as those described in Section~\ref{sec:J1614photo}. We divided the sky-subtracted light curves by the sum of three brighter comparison stars in the field: SDSS~J222904.91+362454.1 ($g=15.4$ mag), SDSS~J222859.80+362532.3 ($g=15.8$ mag), and SDSS~J222902.31+362351.5 ($g=16.2$ mag).

We first observed J2228 over three consecutive nights in 2011 November, but the star went behind the Sun before we could confirm variability. All told, we obtained more than 25.7 hr of photometric observations spread over nearly 10 months, as outlined in Table~\ref{tab:jour}.

\begin{figure}
\centering{\includegraphics[width=0.99\columnwidth]{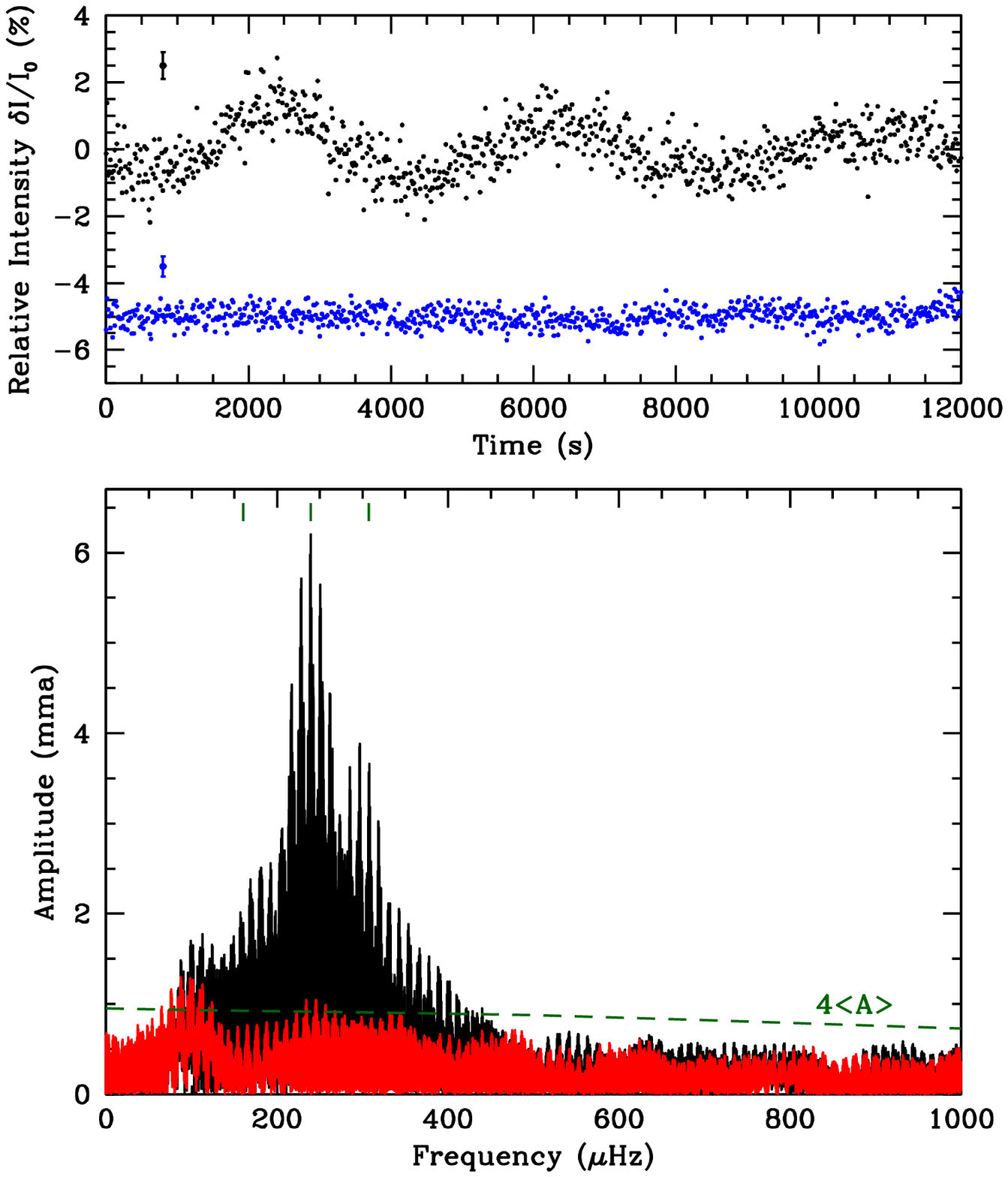}}
\caption{The top panel shows high-speed photometry of J2228 from a representative run on 2012 Sep 20. The brightest comparison star is shown in blue, offset by $-5$\%. Average point-by-point photometric errors are also shown. The bottom panel shows a Fourier transform of our entire data set to date, some 25.7 hr of observations from 2011~November to 2012~September. We also display in red the FT of the residuals after prewhitening by the highest-amplitude periods listed in Table~\ref{tab:J2228freq} and mark those periods with green tick marks at the top of the panel. We mark the 4$<$A$>$ significance level, described in the text, as a dashed green line. \label{fig:J2228}}
\end{figure}

The top panel of Figure~\ref{fig:J2228} shows a portion of a typical light curve for J2228, obtained on 2012 Sep 20, and includes the brightest comparison star in the field over the same run. The bottom panel of this figure shows an FT utilizing all $9327$ light curve points collected thus far. We display the 4$\langle {\rm A}\rangle$ significance line at the bottom of Figure~\ref{fig:J2228}, calculated from the average amplitude of an FT within a 1000 \muhz\ region in steps of 200 \muhz, after pre-whitening by the three highest-amplitude periodicities identified in Family 1 of Table~\ref{tab:J2228freq}.

We identify these periods by taking an initial FT of the data. We iteratively pre-whiten by the highest amplitude peak and take an FT of the residuals, until there are no peaks above our running 4$\langle {\rm A}\rangle$ significance line. As before, the cited errors are not formal least-squares errors to the data but rather the product of $10^5$ Monte Carlo simulations, and the S/N calculation is performed identically to that in J1614.

\begin{table}
 \centering
  \caption{Frequency solutions for SDSS J222859.93+362359.6 \label{tab:J2228freq}}
  \begin{tabular}{@{}lcccc@{}}
  \hline
  ID & Period & Frequency & Amplitude & S/N \\
   & (s) & ($\mu$Hz) & (mma) &  \\
\hline
\multicolumn{5}{c}{\bf Family I: All Data} \\
$f_1$ & $4178.3\pm2.8$ & $239.33\pm0.16$ & $6.26\pm0.14$ & 18.3 \\
$f_2$ & $3254.5\pm2.1$ & $307.27\pm0.20$ & $2.34\pm0.14$ & 7.1 \\
$f_3$ & $6234.9\pm6.0$ & $160.39\pm0.15$ & $1.94\pm0.23$ & 5.4 \\
\hline
\multicolumn{5}{c}{\bf Family II: Only 2012 Sep Data} \\
$f_1$ & $4178.65\pm0.62$ & $239.312\pm0.036$ & $6.44\pm0.20$ & 15.4 \\
$f_2$ & $3254\pm195$ & $307\pm18$ & $2.67\pm0.57$ & 6.7 \\
$f_3$ & $6239\pm1017$ & $160\pm26$ & $2.14\pm0.40$ & 5.1 \\
\hline
\end{tabular}
\end{table}

Because our coverage is so sparse over nearly 10 months, we have computed two families of frequency solutions in Table~\ref{tab:J2228freq}. Family I comes from our entire data set, spanning 2011~November to 2012~September, and is the set of periods that have been pre-whitened to display the red residual FT in Figure~\ref{fig:J2228}. Family II uses only our 2012 September data, 16.6 hr of coverage in good conditions over five nights, and has a considerably cleaner spectral window. Both solutions are in good agreement, although the uncertainties for $f_2$ and $f_3$ in Family II are much larger.

There is also evidence for a formally significant peak at $10{,}075$ s (2.8 hr). However, this periodicity is close to the length of a typical run on this object, and may be an artifact of changing atmospheric conditions, especially differential transparency variations. We have tested this hypothesis by reducing a star with a similar magnitude to J2228 in the field, SDSS~J222901.52+362426.5 ($g=16.9$); one formally significant peak shows up in the FT of that star at a similarly long period, 8999 s. We therefore do not adopt any periods longer than 8900 s in our formal frequency solution. There are no other significant periodicities in the FT of this nearby, similarly bright comparison star.



\section{Discussion}
\label{sec:end}

\subsection{Properties of the First Pulsating ELM WDs}
\label{sec:ensemble}

We can begin to put the first five pulsating putatively He-core ELM WDs into context with the 160 previously known ZZ Ceti stars by exploring the observed properties of both.

Previous studies of the known ZZ Ceti stars have shown convincingly an observed increase in the periods of excited modes with lower effective temperatures \citep{Clemens93,Mukadam06}. This is an expected consequence of cooler ZZ Ceti stars having deeper convection zones, which in turn lengthens the thermal timescale, most important for driving mode instabilities. Empirically, \citet{Mukadam06} showed there was a roughly linear increase in the weighted mean period (WMP) of ZZ Ceti stars with decreasing effective temperature.

This trend generally holds true with the pulsating ELM WDs, as well, which have significantly cooler temperatures and longer periods than their CO-core brethren. Table~\ref{tab:5props} shows the WMPs of the known pulsating ELM WDs, as well as the range of periods observed, which extends up to 6235 s in J2228, the coolest DAV known to date. The longer periods also make sense in the context of these WDs having lower surface gravities (and thus lower mean densities), since the period of pulsation modes roughly scales with the dynamical timescale for the whole star, $\Pi \propto \rho^{-1/2}$.

Two independent groups have recently published low-mass WD models, and both predict both $g$- and $p$-mode pulsational instabilities in ELM WDs. For the $g$-mode pulsations, \citet{Corsico12} found that only higher-radial-order $\ell=1$ ($k \geq 9$) were unstable, and thus they predict $\ell=1$ pulsation periods $\Pi > 1100$ s. This is consistent with the observed distribution of long-period variability in the pulsating ELM white dwarfs discovered so far (Table~\ref{tab:5props}). Similarly, \citet{VanGrootel13} found that only $k \geq 4$ $g$-modes were unstable in their non-adiabatic calculations, suggesting mode periods in excess of $\Pi > 500$ s.

While we do not yet have a sufficient suite of He-core WD models to match against for a full asteroseismic analysis, we can infer some early conclusions about the physical nature of our first five pulsating ELM WDs. For example, it is possible that $f_1$ and $f_2$ of J1614, discussed in Section~\ref{sec:J1614photo}, are consecutive radial modes. If true, then the difference in their periods, $78.56$ s, could probe the forward mean period spacing of J1614, which is a sensitive function of the overall WD mass. We note that \citealt{Corsico12} find that low-mass WDs reach the asymptotic limit for mean period spacing for only high radial order, $k>25$. For reference, they find an asymptotic mean period of 94.3 s for a 0.20 \msun, 8860 K He-core WD, roughly the values we derive from the spectroscopy.

The models of \citet{Corsico12} provide a useful context for the observed periodicities in J1614: Their 8889 K, 0.22 \msun\ He-core WD model (which has a relatively thick hydrogen layer, $M_{\rm H}/M_* = 10^{-2.78}$), shows an $\ell=1,k=11$ $g$-mode at 1196.07 s and an $\ell=1,k=12$ $g$-mode at 1274.67 s. In addition, we find a good match of these two periods to their 8850 K, 0.303 \msun\ model, in which the $\ell=1,k=14$ and $\ell=1,k=15$ $g$-modes occur at 1196.07 s and 1274.66 s, respectively, and differ by $78.60$ s. However, seismology is made extremely difficult by the detection of just two periodicities in J1614.

It is also possible that the 52.5 \muhz\ difference between these two oscillations could be explained by a rotational splitting from a single mode in J1614 \citep{Hansen77}. Such a splitting could arise from a 16.9 hr rotation rate if the 1184 s mode is an $\ell=1$ mode, assuming roughly solid-body rotation.

\begin{table}
 \centering
  \caption{Properties of the Five Known Pulsating ELM WDs  \label{tab:5props}}
  \begin{tabular}{@{}llll@{}}
  \hline
  Property & Value & Property & Value \\
 \hline
\multicolumn{4}{c}{\bf SDSS~J184037.78+642312.3} \\
\teff\  & $9390\pm140$ K    		& \logg\   	& $6.49\pm0.06$ \\
Mass    & $\sim$0.17 \msun\ 		& $g$-band	& 18.8 mag \\
Periods & $2094-4890$ s     		& WMP     	& $3722$ s \\
\multicolumn{4}{c}{\bf SDSS~J111215.82+111745.0} \\
\teff\  & $9590\pm140$ K    		& \logg\   	& $6.36\pm0.06$ \\
Mass    & $\sim$0.17 \msun\ 		& $g$-band	& 16.2 mag \\
Periods & $107.6-2855$ s    		& WMP	  	& $2288$ s \\
\multicolumn{4}{c}{\bf SDSS~J151826.68+065813.2} \\
\teff\  & $9900\pm140$ K    		& \logg\   	& $6.80\pm0.05$ \\
Mass    & $\sim$0.23 \msun\ 		& $g$-band 	& 17.5 mag \\
Periods & $1335-3848$ s     		& WMP	  	& $2404$ s \\
\multicolumn{4}{c}{\bf SDSS~J161431.28+191219.4} \\
\teff\  & $8800\pm170$ K    		& \logg\   	& $6.66\pm0.14$ \\
Mass    & $\sim$0.19 \msun\ 		& $g$-band 	& 16.4 mag \\
Periods & $1184-1263$ s     		& WMP	  	& $1235$ s \\
\multicolumn{4}{c}{\bf SDSS~J222859.93+362359.6} \\
\teff\  & $7870\pm120$ K    		& \logg\   	& $6.03\pm0.08$ \\
Mass    & $\sim$0.16 \msun\ 		& $g$-band 	& 16.9 mag \\
Periods & $3254-6235$ s     		& WMP	  	& $4958$ s \\
\hline
\end{tabular}
\end{table}

\subsection{The \logg--\teff\ Diagram}
\label{sec:rededge}

\begin{figure*}
\centering{\includegraphics[width=0.85\textwidth]{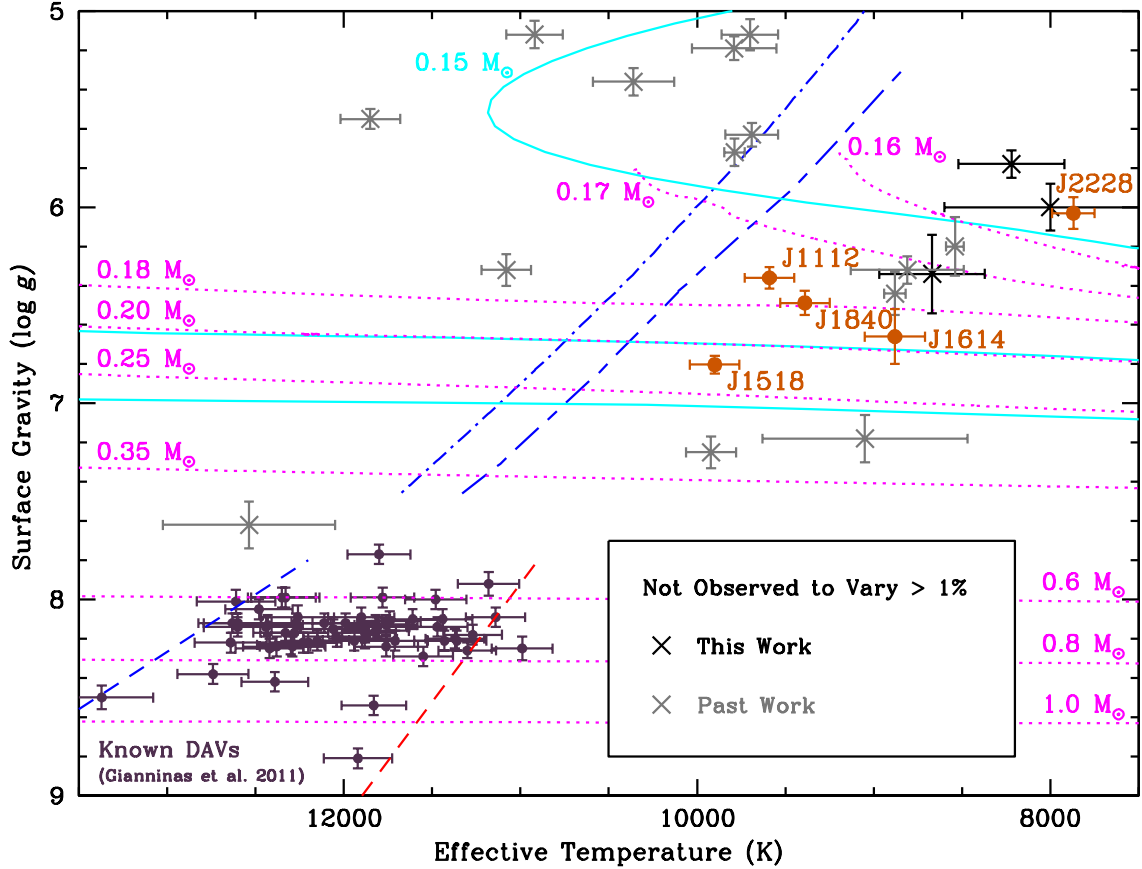}}
\caption{The \logg--\teff\ diagram for pulsating DA WDs. We show 56 CO-core ZZ Ceti stars characterized in a consistent way by \citet{Gianninas11} as purple dots, and mark the five known pulsating ELM WDs in burnt orange. We denote an extrapolated theoretical blue edge for the low-mass WD instability strip; this dashed-dotted blue line is described in the text. We also include as a long-dashed-short-dashed blue line the theoretical blue edge for low-mass ZZ Ceti stars from \citet{VanGrootel13}. We mark the empirical blue- and red-edges for CO-core ZZ Ceti stars from \citet{Gianninas11} as dashed blue and red lines, respectively. Objects not observed to vary to larger than 10 mma (1\%) are marked with an X. We include three new WDs not observed to vary, listed in Table~\ref{tab:null}; the others were detailed in \citet{Steinfadt12,Hermes12,Hermes13}. Cooling models for different WD masses are included as dotted and solid lines and described in the text. \label{fig:search}}
\end{figure*}

\begin{table*}
 \centering
  \caption{Newly Observed Low-Mass DAV Candidates and Null Results\label{tab:null}}
  \begin{tabular}{@{}lccccc@{}}
  \hline
  Object & $g$-SDSS & \teff\ & \logg\ & Reference & Det. Limit \\
   & (mag) & (K) & (cm s$^{-1}$) &  & (\%)  \\
 \hline
SDSS~J070216.21+111009.0 & 16.1 & $8800\pm600$  & $6.00\pm0.12$ & \citet{BrownELMiii} & 0.3 \\
SDSS~J090052.04+023413.8 & 18.0 & $8220\pm300$  & $5.78\pm0.07$ & \citet{BrownELMiii} & 0.4 \\
PSR~J101233.42+530702.8 & 19.6 & $8670\pm300$  & $6.34\pm0.20$ & \citet{Callanan98} & 0.7 \\
\hline
\end{tabular}
\end{table*}

We may compare the first five pulsating ELM WDs to the previously known ZZ Ceti stars by placing them in a \logg--\teff\ diagram, shown in Figure~\ref{fig:search}. Doing so, we discover there are at least six ELM WDs with temperatures and surface gravities between the newfound pulsating ELM WD J2228 and the other four known pulsating ELM WDs. These non-variable ELM WDs have been observed extensively and do not show significant evidence of pulsations to at least 1\% relative amplitude. We have excellent limits on the lack of variability in four of these six, ruling out pulsations larger than 0.3\% amplitude.

We have put limits on three of these new non-detections, detailed in Table~\ref{tab:null}. We note that \citet{Steinfadt12} previously observed PSR~1012+5307, but we have put much more stringent limits on a lack of variability on this faint ELM WD with 7 hr of observations in excellent conditions.

The other three interlopers have been detailed in previous studies. SDSS~J0822+2753 is a \teff\ $=8880\pm60$ K, \logg\ $=6.44\pm0.11$ WD observed not to vary to 0.2\% \citep{Hermes12}. SDSS~J1443+1509 is a \teff\ $=8810\pm320$ K, \logg\ $=6.32\pm0.07$ WD with exquisite limits on lack of variability, to $< 0.1$\% \citep{Hermes13}. Finally, NLTT~11748 is the \teff\ $=8540\pm50$ K, \logg\ $=6.20\pm0.15$ primary WD in an eclipsing WD+WD binary \citep{Steinfadt10NLTT}. It was shown by \citet{Steinfadt12} not to vary out of eclipse to above 0.5\%. We have obtained an additional 8 hr of photometry of NLTT~11748 out of eclipse at McDonald Observatory and can independently rule out variability larger than 0.3\%.

The discovery of pulsations in J2228, which is considerably cooler than at least a half-dozen other photometrically constant ELM WDs, questions the purity of the instability strip for He-core WDs and confuses the location of an empirical red edge. However, there is no a priori reason to expect the ELM WD instability strip to be pure, or for there to exist a connected low-mass extension of the classical CO-core ZZ Ceti instability strip; evolution through a specific temperature-gravity region is not well established for the ELM WDs, and they may not all cool through the instability strip in as simple a manner as the CO-core ZZ Ceti stars. In fact, some of these ELM WDs may indeed be in the throws of unstable hydrogen shell burning episodes; they may not be cooling at all, but rather looping through the HR diagram prior to settling on a final cooling track (e.g., \citealt{Althaus13}). Such excursions are not expected for CO-core ZZ Ceti stars, which are expected to monotonically cool through an observationally pure instability strip.

We have plotted the evolution of theoretical cooling tracks for several different WD masses through the effective temperatures and surface gravities in Figure~\ref{fig:search}. We plot the 0.16 \msun, 0.17 \msun, 0.18 \msun, 0.20 \msun, 0.25 \msun, and 0.35 \msun\ He-core models of \citet{Panei07} as dotted magenta lines. We have also used the stellar evolution code MESA \citep{Paxton11,Paxton13} to model the evolution of 0.15 \msun, 0.20 \msun, and 0.25 \msun\ He-core WDs, shown as solid cyan lines in Figure~\ref{fig:search}. For reference, we have also included 0.6 \msun, 0.8 \msun, and 1.0 \msun\ CO-core cooling tracks \citep{HB06,KS06,Tremblay11}.

Where the lowest-mass WD models enter this diagram depends on how we artificially remove mass from the models, and there is a very noticeable discrepancy between the 0.16 \msun\ \citet{Panei07} WD models and our 0.15 \msun\ WD models using MESA. As an added complication, except for the lowest-mass ELM WDs (below roughly 0.18 \msun), recurrent hydrogen shell flashes cause the ELM WD model to loop many times through this \teff-\logg\ plane, further confusing the picture \citep{Panei07,Steinfadt10,Althaus13}. Thus, it is not entirely surprising to find non-variable ELM WDs between J2228 and the four warmer pulsating ELM WDs. Further empirical exploration of the entire ELM WD instability strip offers a unique opportunity to constrain physical and evolution models of ELM WDs, specifically these late thermal pulses and the mass boundary for the occurrence of these episodes. Long-term monitoring of the rate of period change of pulsating ELM WDs also affords an opportunity to constrain the cooling (or heating) rate of these objects \citep{WinKep08}.

In contrast to the confusion along the red edge of the instability strip, the blue edge is more reliably predicted by theory. The theoretical blue edge (dotted blue line) in Figure~\ref{fig:search} has been calculated following \citet{Brickhill91} and \citet{Goldreich99}. We use the criterion that $P_{\rm max} \sim 2 \pi \tau_{\rm C}$ for the longest period mode that is excited, where $P_{\rm max}$ is the mode period and the timescale $\tau_{\rm C}$ describes the heat capacity of the convection zone as a function of the local photospheric flux, which we compute from a grid of models (see \citealt{Hermes13} for further details). We use the criterion $2 \pi \tau_C = 100$ s, with the convective prescription ML2/$\alpha$=1.5. We also include the theoretical blue edge of \citet{VanGrootel13}, which uses a slightly less efficient prescription for convection, ML2/$\alpha$=1.0.



\section{Conclusions}

We have discovered pulsations in two new extremely low-mass, putatively He-core WDs using optical facilities at the McDonald Observatory. Spectral fits show that these two ELM WDs, J1614 and J2228, are the coolest pulsating WDs ever found. This brings to five the total number of pulsating ELM WDs known, establishing them as a new class of pulsating WD. As with the more than 160 CO-core ZZ Ceti stars that have been known for more than four decades, the luminosity variations in these ELM WDs is so far consistent with surface temperature variations caused by non-radial $g$-mode pulsations driven to observability by a hydrogen partial ionization zone.

The coolest pulsating ELM WD, J2228, has a considerably lower effective temperature than six similar-gravity ELM WDs that are photometrically constant to good limits. In contrast to the CO-core ZZ Ceti stars, which are believed to represent a stage in the evolution of all such WDs, ELM WDs may not all evolve through an instability strip in the same way, and thus we may not observe their instability strip to be pure. Theoretical He-core WD models predict multiple unstable hydrogen-burning episodes, which complicates the evolution of an ELM WD through a simple instability strip. Empirically discovering ELM WDs in this space that do or do not pulsate opens the possibility to use the presence of pulsations in ELM WDs to constrain the binary and stellar evolution models used for low-mass WDs, which may better constrain these poorly understood CNO-flashing episodes.


\section*{Acknowledgments}

We acknowledge the anonymous referee for valuable suggestions that greatly improved this manuscript. J.J.H., M.H.M. and D.E.W. acknowledge the support of the NSF under grant AST-0909107 and the Norman Hackerman Advanced Research Program under grant 003658-0252-2009. M.H.M. additionally acknowledges the support of NASA under grant NNX12AC96G. B.G.C. thanks the support from CNPq and FAPERGS-Pronex-Brazil. The authors are grateful to the essential assistance of the McDonald Observatory support staff, especially Dave Doss and John Kuehne, and to Fergal Mullally for developing some of the data analysis pipeline used here.

\label{lastpage}

\end{document}